# Enhanced Solar Water Splitting by Swift Charge Separation in Au/FeOOH Sandwiched Single Crystalline Fe$_2$O$_3$ Nanoflake Photoelectrodes

Lei Wang,[a] Nhat Truong Nguyen,[b] Yajun Zhang,[a] Yingpu Bi,*[a] and Patrik Schmuki*[b,c]

[a]  Prof. Dr. Lei Wang, Dr. Yajun Zhang, Prof. Dr. Yingpu Bi
State Key Laboratory for Oxo Synthesis and Selective Oxidation, National Engineering Research Center for Fine Petrochemical Intermediates, Lanzhou Institute of Chemical Physics, CAS, 730000 Lanzhou, China.
E-mail: yingpubi@licp.cas.cn

[b]  Nhat Truong Nguyen, Prof. Dr. Patrik Schmuki
Department of Materials Science and Engineering, WW4-LKO, University of Erlangen-Nuremberg, Martensstrasse 7, D-91058 Erlangen, Germany.
E-mail: schmuki@ww.uni-erlangen.de

[c]  Prof. Dr. Patrik Schmuki
Department of Chemistry, King Abdulaziz University, 80203 Jeddah, Saudi Arabia Kingdom.

**Abstract**: In this work, single crystalline α-$Fe_2O_3$ nanoflakes (NFs) are formed in a highly dense array by Au seeding of a Fe substrate by a thermal oxidation technique. The NFs are conformally decorated with a thin FeOOH cocatalyst layer. Photoelectrochemical (PEC) measurements show that this photoanode with the α-$Fe_2O_3$/FeOOH NFs rooted on the Au/Fe structure exhibits a significantly enhanced PEC water oxidation performance compared to the plain α-$Fe_2O_3$ nanostructure on the Fe substrate. The α-$Fe_2O_3$/FeOOH NFs on Au/Fe photoanode yields a photocurrent density of 3.1 mA $cm^{-2}$ at 1.5 $V_{RHE}$, and a remarkably low onset potential of 0.5-0.6 $V_{RHE}$ in 1 M KOH under AM 1.5G (100 mW $cm^{-2}$) simulated sunlight illumination. The enhancement in PEC performance can be attributed to a synergistic effect of the FeOOH top decoration and Au under-layer. While FeOOH facilitates hole transfer at the interface of electrode/electrolyte, the Au layer provides a sink for the electron transport to the back contact: this leads overall to a drastically improved charge-separation efficiency in the single crystalline α-$Fe_2O_3$ NF photoanode.

Photoelectrochemical (PEC) water splitting to produce hydrogen fuel with abundant solar energy is a promising strategy to resolve the current energy and environmental crisis.[1-3] Among various semiconductors, hematite (α-$Fe_2O_3$), has recently re-emerged as a promising photoanode material for water-splitting owing to its favorable band gap (~2.1 eV), high natural abundance, low cost, and environmental friendliness. However, the relatively low conversion efficiencies owing to an intrinsically poor electrical conductivity, a short hole diffusion length (2-4 nm), and a sluggish water oxygen kinetics, greatly restrict its application.[4-7] Moreover, $Fe_2O_3$ photoanodes composed of nanoparticles, usually suffer from charge migration and recombination problems at or in grain boundaries. Recently, photoanodes comprised of one-dimensional (1D) nanoarrays (e.g. nanorods and nanotubes) grown vertically to the conductive substrate (e.g. fluorine-doped tin oxde (FTO)) have become of particular interests due to a direct electron transport pathway and short hole diffusion length to the interface, which result in improved PEC properties compared to traditional particle/porous films.[7-11] More recently, vertically aligned nanosheet arrays with specifically designed different crystal facets were reported. [12] Due to the different relative energies of different facets, photoexicited electrons and holes can be driven to different crystal facets, and thus certain facets of a semiconductor provide reduction while others facilitate oxidation.[12c] Thus a suitable combination exposed crystal facets should greatly decrease the recombination ratios of electron-hole pairs.

Except for faceting, reducing surface-mediated charge recombination can be achieved by passivating surface electron trap states by thin overlayer coatings.[13] Furthermore, increasing the rate constants for holes transfer from the photoelectrode to molecular reactants can be obtained by surface modification with oxygen evolution catalysts (OEC), that are capable of transferring photogenerated holes to $H_2O$ or $OH^-$ more rapidly, and thus reducing the water oxidation overpotential. Early reports on ruthenium oxide ($RuO_2$) and iridium oxide ($IrO_2$) as cocatalysts with α-$Fe_2O_3$ have demonstrated to significantly improve the oxygen evolution reaction (OER) performance.[14-16] Noble metal-free OER cocatalysts, e.g. cobalt-based materials ($Co_3O_4$, Co-Pi et al.), also exhibit a promotive effect on the activity of α-$Fe_2O_3$ for PEC water oxidation leading to an enhanced photocurrent density and a shift of the oxygen evolution onset potential.[17-21] Recently oxyhydroxides (FeOOH, NiOOH and FeNi layered double hydroxide) have been developed as highly cxefficient cocatalysts for semiconductor photoanodes to improve the OER process.[22-27] For example, Ye et al.[22] reported that α-$Fe_2O_3$ photoanodes decorated with FeOOH nanospikes produced by photodeposition showed an enhancement of the photocurrent density compared to a bare α-$Fe_2O_3$ photoelectrode, due to a significant co-catalytic effect of FeOOH. Considering the present state of α-$Fe_2O_3$ photoanodes, still the construction of electrode configurations that can only provide an ideal light absorption but also optimize carrier transport and transfer, to achieve not only a high conversion efficiency but also a low onset potential, are of particular importance, and remain a great challenge.

In this work, we first show that a gold nanoparticle (Au NP) seed layer on an iron substrate can be used to trigger the thermal growth of a high density array of single crystalline α-$Fe_2O_3$ nanoflakes (NFs). The Au NPs remain as an under-layer between the NFs and the iron substrate. The single crystalline α-$Fe_2O_3$ NFs then can be decorated with a FeOOH cocatalyst by a precipitation method, without any change of crystallographic structure and/or morphology. The Au at the NFs/Fe interface provides an electron sink, and thus an ideal junction to the conducting surface. The synergistic effect between FeOOH as an OEC and the Au sublayer not only significantly facilitates charge transfer at the surface, but also increases the electron transport to the back contact. As a result, a considerably enhanced water splitting performance of such photoanodes is obtained with a photocurrent density of 3.1 mA cm$^{-2}$ at 1.5 V$_{RHE}$, and an onset potential of 0.5-0.6 V$_{RHE}$ (with AM 1.5G simulated sunlight, 100 mW cm$^{-2}$, in 1 M KOH). This is the lowest onset value combined with the high photocurrent for PEC water oxidation on a non-doped 1D α-Fe$_2$O$_3$ photoanode. Moreover, to the best of our knowledge, this is to achieve a bimodal separation of electron-hole pairs through a single crystalline hematite layer without using additional element doping. Furthermore, a synchronous light X-ray photoelectron spectroscopy (SIXPS) technique is used to elucidate the mechanism of FeOOH OEC on the α-Fe2O3 photoanodes.

Scheme 1 illustrates the synthesis process of the α-Fe$_2$O$_3$ NFs formed on the Fe substrate. First, an Au NPs layer was seeded on Fe foil by a self-induced decoration (Figure S1). For this, the Fe foil was immersed in a HAuCl$_4$ aqueous solution for a defined time, leading to Au NPs layer with a typical size of 5-20 nm by galvanic displacement and the formation of FeOOH initiation seeds – a more detailed description of this process that creates a high density of initiation sites for the thermal growth of Fe$_2$O$_3$ 1D structures is given in the SI (see Figure S2). After thermal oxidation of Au/Fe foil (400 °C for 2 h in air), the surface is covered with a dense and ordered array of 1D α-Fe$_2$O$_3$ NFs (Figure 1a). By contrast, without the formation of Au NPs layer on the substrate, the as-grown sample shows a much lower density of NFs on the top surface (Figure S3).

In order to decorate the NFs with the FeOOH cocatalyst, the as-grown NF arrays are immersed in a FeCl$_3$ aqueous solution from 1 h to 45 h by a facile "top-down" method (see SI). Scanning electron microscopy (SEM) was performed to characterize the morphology change of α-Fe$_2$O$_3$ NFs photoelectrodes without and with the Au NPs layer (Figures S3-S4, and Figure 1a-d). In Figure 1a and c, it is apparent that the α-Fe$_2$O$_3$ NFs on Au/Fe foil form a dense array with individual flakes with a sharp apex. The flakes are 1.5-2.5 μm in length, 100-400 nm thick at the base, and the thickness tapers down to ~10 nm at the tip. The Au remains accumulated as an Au NP layer located at the interface between a thermal magnetite (Fe$_3$O$_4$) layer and the iron substrate (Figure 1c and d). SEM-EDS mapping images (Figure 1d and Figure S5) clearly show an Au signal that spatially correlates with the Fe and O from the interface of oxide layer/substrate. Figure 1e-i show high resolution transmission electron microscopy (HRTEM) images and SAD patterns of single α-Fe$_2$O$_3$ NFs formed on Au/Fe layer. The clear lattice images (Figure e and f) indicate the high crystallinity and the single crystalline nature of the α-Fe$_2$O$_3$ NFs. A lattice spacing of 0.25 nm for the [110] lattice planes of the α-Fe$_2$O$_3$ structure along the NFs (c axis) can be readily resolved. (For comparison, the α-Fe$_2$O$_3$ NFs on Fe substrate in Figure S6 show a relatively lower crystallinity.) Figure 1j and k show the model of α-Fe$_2$O$_3$ crystal structure and surface atom arrangement of a [110] crystal plane, illustrating the high density of iron atoms on this plane [110].[28] For these planes, conductivities have been reported up to 4 orders of magnitude higher for the basal plane than in the perpendicular direction.[2,29] After further FeCl3 treatment (Figure 1b and Figure S7), the samples have a similar morphology to that of α-Fe$_2$O$_3$ NFs on the Au/Fe foil (Figure 1a). The HRTEM image and SAD pattern clearly show an amorphous coating layer of ~2 nm on the NFs (Figure 1f and i).

The decorated NFs were then investigated for PEC performance under AM 1.5G sun irradiation in 1 M KOH electrolyte. Figure 2a shows the transient photocurrent-potentiacurves of α-Fe2O3 NFs on Fe, α-Fe$_2$O$_3$ NFs on Au/Fe, and α-Fe$_2$O$_3$ /FeOOH NFs on Au/Fe. The optimization of Au deposition and annealing treatment are shown in Figures S9-S11. In Figure 2a, upon sweeping the potential from 0.4 to 1.7 V$_{RHE}$ under AM 1.5G illumination, the as-grown α-Fe$_2$O$_3$ NF sample on Fe foil annealing at 400 °C for 2 h exhibits a water-oxidation onset potential of 0.5 V$_{RHE}$ (Figure S8), and a photocurrent density increases of 0.22 mA cm$^{-2}$ at 1.23 V$^{RHE}$. For the α-Fe$_2$O$_3$ NFs on Au/Fe immersion in HAuCl$_4$ solution for 1 min, the photoresponse over the whole potential range from 0.5 to 1.6 $V_{RHE}$ is increased, and the photocurrent density is up to 0.64 mA cm$^{-2}$ at 1.23 $V_{RHE}$. In Figure S9, as the deposition time of Au is increased up to 30 min or 60 min, the photocurrent decreases obviously compared to the samples with 1 min and 10 sec. deposition, which is due to the higher amount of Au particles that exist in the inside and on the top surface (Figure S10g and h). This is reflected in a hampered water oxidation kinetic in comparison to the samples treated for a short deposition time. In Figure 2a upon further FeOOH cocatalyst treatment (in FeCl$_3$ solution for 28 h) on the α-Fe$_2$O$_3$ NFs on Au/Fe, the photocurrent density shows a significant enhancement, especially at 1.23 $V_{RHE}$ up to 1.45 mA cm$^{-2}$, and a maximum photocurrent density of 3.1 mA cm$^{-2}$ is obtained at 1.5 $V_{RHE}$. The solar energy conversion efficiency of corresponding sample is calculated as 0.23% at 0.8 $V_{RHE}$ (Figure 2b). The onset potential (inset of Figure 2a) for the NFs structure does not shift in the cathodic direction after cocatalyst decoration, whereas it shows a higher photocurrent at 0.6-0.7 $V_{RHE}$ compared to those of α-Fe$_2$O$_3$ NFs on the Fe or Au/Fe substrate.

Figure 2c shows the results of incident photocurrent efficiency (IPCE) versus incident light wavelength for the photoanodes, measured at 1.5 $V_{RHE}$ in 1 M KOH. The IPCE values are in line with the measurements using AM 1.5G sun light irradiation (Figure 2a). Over the entire spectrum range from 300 nm to 600 nm, the α-Fe$_2$O$_3$ NFs on Au/Fe sample shows an improved IPCE value compared to the α-Fe$_2$O$_3$ NFs on Fe, and the introduction of FeOOH cocatalyst on the NFs leads to a significantly enhanced IPCE value compared to the FeOOH-free electrode. The maximum IPCE value at 400 nm is 40% for the α-Fe$_2$O$_3$/FeOOH NFs on Au/Fe. The enhancement of the photoresponse can be ascribed to the cocatalyst quickly transferring the hole from the NFs to the electrolyte, and the increased electrical contacts between NFs/magnetite and the substrate (aiding the transfer of the electron to the back contact). Stability measurements of the corresponding α-Fe$_2$O$_3$ NF electrodes are shown in Figure 2d, in which the photocurrents of the samples keep stable values up to 100 min irradiation. The fluctuation of photocurrents is due to the O$_2$ released from the electrodes observed during the water splitting reaction. The ultraviolet-visible (UV-vis) absorbance spectra were further examined to characterize for the photoactivity of photoanodes, as shown in Figure 2e. The α-Fe$_2$O$_3$ NFs on Au/Fe sample shows a higher absorption than α-Fe$_2$O$_3$ NFs on Fe, whereas for the α-Fe$_2$O$_3$/FeOOH NFs on Au/Fe, a lower absorbance compared to the non-treated FeOOH sample is obtained. The latter is likely due to partly iron corrosion happened in the acid solution. The resulting Eg values are in the order of 2.1 eV (Figure 2f), in line with reported data for hematite.[1,3]

Furthermore, various decoration times of FeOOH cocatalyst on α-Fe$_2$O$_3$ NFs of Au/Fe were examined; the solar water splitting performance is shown in Figure S12. From Figure S12a and b one can deduce that after a shorter immersion time (1~4 h), the photocurrent densities decrease compared to the as-grown electrode on Au/Fe. This decrease may be due to an etching of the nanoflakes after immersion in acidic solution, only after a longer immersion time (>8 h, Figure S12c-h), the electrodes become significantly coated with FeOOH and thus exhibit a superior PEC performance. In line with literature,[24] an amorphous FeOOH layer can improve the water oxidation kinetics and passivate surface states of hematite. Moreover, the intimate and energetically favorable contact between α-Fe$_2$O$_3$ and FeOOH leads to a reduced recombination rate of electron-hole pairs at the α-Fe$_2$O$_3$/FeOOH interface.[25,27] A further extension of the immersion time up to 45 h (Figure S12i) causes a decay in the photoresponse. This indicates hampered water oxidation kinetics in comparison to the electrodes after 8-40 h treatment, most likely due to the decay of nanostructure as observed in SEM image of Figure S7i.

Figure 3a shows X-ray diffraction (XRD) patterns of α-Fe$_2$O$_3$ NFs on Fe or Au/Fe, and α-Fe$_2$O$_3$/FeOOH NFs on Au/Fe. The Au NPs layer on Fe substrate after self-reduction is also shown for comparison. The intense peaks at 2θ = 38.2°, 44.7°, 64.9°, and 77.8° can be assigned to Au reflexes.[29] For all NF electrodes, hematite and magnetite signals are evident, and for the flakes on the Au/Fe, Au diffraction peaks (38.2°, 44.6°, and 65.1°) are detected. This corresponds to the SEM observation in Figure 1c and d. No obvious change in the XRD pattern for the NFs with the treatment of FeOOH is observed compared to non-treated one. To gain some insight into the nature of the amorphous regions, we characterize the samples using Raman spectroscopy. In Figure 3b, five of seven possible optical modes (2A$_g$+5E$_g$) for the NF samples are observed with correspondence: A$_g$, 223 and 497 cm$^{-1}$, and E$_g$, 289, 408, and 610 cm$^{-1}$. The extra peak centered at 665 cm$^{-1}$ can be assigned to magnetite, which usually exists in hematite samples prepared by thermal oxidation.[30] The relatively weak Raman signal for the α-Fe$_2$O$_3$ NFs on Fe can be attributed to the low crystallinity, as confirmed by the corresponding low diffraction intensity in the XRD patterns (Figure 3a). However, no additional peaks assigned to the FeOOH can be observed. In a reference sample, we deposited the FeOOH layer directly on the Au/Fe substrate. The resulting spectrum is shown in the inset of Figure 3b where broad peaks at 231, 549, 692 cm$^{-1}$ are observed, with positions corresponding to those found for FeOOH.[29] The fourier transform infrared (FTIR) spectrum of the α-Fe$_2$O$_3$/FeOOH NFs on Au/Fe (Figure 3c) matches well with the standard spectrum observed in the corresponding bulk.[31] The bands at 1600, 1391, 1367, 1005, and 830 cm$^{-1}$ can be ascribed to the Fe-O vibrational modes in FeOOH.

XPS shows no trace of Cl (Figure S13), indicating that Cl concentration is below the detection level of analyses (below ~1at%). In the Fe 2p region (Figure S14), the Fe 2p spectra show a binding energy (BE) of 724.8 eV (Fe 2p$_{1/2}$) and 710.9 eV (Fe 2p$_{3/2}$) with a shake-up satellite line at 719.3 eV. In Figure 3e, the lowest BE peak of O 1s at 529.9 eV is attributed to oxygen atoms in the iron oxide lattice, and the peak at 531.5 eV is assigned to the hydroxyl group.

The photoinduced charge transfer and separation process plays a crucial role in reaching high light-to-energy conversion efficiencies. Here we use a XPS technique that is combined with synchronous light illumination (SIXPS) to directly extract charge transfer information (Figure 3d). Under light illumination, the peaks were detected to shift resulting from the electron density changes of various atoms in the excited state, which would directly reveal the separation efficiency and transfer direction of photoexcited charges.[32] Figure 3e-g show the results from this technique. Furthermore, under light illumination, distinct changes on the BE shift of O 1s peak in the α-Fe$_2$O$_3$ NFs can be observed. More specifically, in the case of α-Fe$_2$O$_3$ NFs on Fe (Figure 3e), no evident shift of all peaks has been detected either under light or in dark. However, for the α-Fe$_2$O$_3$/FeOOH NFs on Au/Fe under light illumination (Figure 3g), the BE of O 1s at 529.9 eV shifts to a lower binding energy of 529.8 eV. A similar effect is also observed in the α-Fe$_2$O$_3$ NFs on Au/Fe (Figure 3f), while its shift is slightly smaller than that of α-Fe$_2$O$_3$/FeOOH NFs on Au/Fe. Moreover, when the α-Fe$_2$O$_3$/FeOOH NFs on Au/Fe immersed in FeCl3 solution for 1 h, no evident shift in the O 1s peaks has been detected (not shown). In contrast, the BE of Fe 2p does not shift after light illumination (Figure S14), which is due to the effect of magnetite in the iron.

These findings are consistent with the PEC performance as shown in Figure 2a and Figure S12. On the basis of above results, it could be concluded that appropriate immersion can facilitate an efficient charge separation in hematite. As reported in literatures,[33,34] the density of the iron atoms in the crystal plane of the [110] surface is 10.1 atoms nm$^{-2}$, i.e., much higher than on other planes, e.g. [001]

plane with an actual value of 2.3 atoms nm$^{-2}$. We consider the [110] the most important crystal plane influencing the PEC activity. Under light illumination, the electrons are excited from the iron orbitals to the oxygen orbitals, the surface termination is crucial (Figure 3f). After the FeOOH OEC decoration on the photoanodes, charge transfer is facilitated - trapped surface charge being evident from a shift of O 1s peak.

Electrochemical impedance spectroscopy (EIS) provides further information on the interfacial properties of the electrodes. Figure 4a shows typical EIS curves for the α-Fe$_2$O$_3$ NFs on Au or Au/Fe, and α-Fe$_2$O$_3$/FeOOH NFs on Au/Fe under illumination, respectively. The α-Fe$_2$O$_3$/FeOOH NFs on Au/Fe show the lowest charge transfer resistance (396 Ω), which means that the FeOOH layer is able to reduce surface trapping states, and facilitate charge transfer at the electrode interface. Water oxidation is thus easier to occur leading to a higher photocurrent as apparent in the I-t curves (Figure 2a). The effect of FeOOH and Au sublayer on enhancing PEC performance of α-Fe$_2$O$_3$ NF photoanodes can be seen in intensity modulated photocurrent spectroscopy (IMPS) as shown in Figure 4b. In the measurement, an ac perturbation of the light intensity is superimposed on illumination of the electrode at an applied potential, and the periodic photocurrent response of the system is reported as a functional of the modulation frequency. The α-Fe$_2$O$_3$ NF samples on Au/Fe without and with FeOOH decoration show more than two decade faster electron transport kinetics compared to the α-Fe$_2$O$_3$ NFs on Fe, and the α-Fe$_2$O$_3$/FeOOH NFs on Au/Fe exhibit the highest electron transport kinetics of all the samples due to a suitable band structure (i.e. junction cascade toward the back contact). Thus the number of charge carriers reaching the back electrode is drastically increased for the α-Fe$_2$O$_3$ NFs on Au/Fe due to the faster transfer, resulting in an increase of the PEC performance. Other contribution to the fast electron transport may be the high crystallinity of α-Fe$_2$O$_3$ NFs on Au/Fe substrate.[35-37]

In line with above results, the existence of FeOOH decreases the reaction barriers for water oxidation and facilitates the holes transfer to the electrode/electrolyte interface, accounting for the significantly improved PEC performance. Meanwhile, the Au sublayer used for seeding high density NF-structures provides a sink effect that aids the electron transfer from the α-Fe$_2$O$_3$/Fe$_3$O$_4$ to the iron substrate. This can be interpreted in terms of a gradient structure Fe$_2$O$_3$/Fe3O4 and the Au under-layer, which can facilitate the interfacial electron transfer from the top α-Fe$_2$O$_3$ photoactive layer to the substrate, in line with a favorable Fermi level sequence. The work function is 5.52 eV for Fe$_3$O$_4$ and 5.1 eV for Au.[38,39] The above findings illustrate that the Au-FeOOH layers-gradients are an efficient strategy to achieve beneficial effects for the transport of electrons and holes. Additionally, for the 1D α-Fe$_2$O$_3$ nanorods grown onto a FTO substrate using a hydrothermal method the photocurrent of α-Fe$_2$O$_3$ nanorods is improved, while the onset potential is still very high (0.8~0.9 V$_{RHE}$), and the relatively higher photoresponse is normally attributed to the Sn or Ti doping.[9,10,36] Thus, this work provides an alternate and more successful strategy for the fabrication of highly efficient semiconductor photoanodes in PEC water oxidation.

In the present work we show how to form high density single crystalline α-Fe$_2$O$_3$ nanoflake arrays on a substrate using Au seeding and thermal oxidation. The α-Fe$_2$O$_3$ NFs can be decorated with a FeOOH cocatalyst by a precipitation method. A synergistic effect between FeOOH and Au not only significantly facilitates charge transfer at the interface of electrode/electrolyte, but also increases the electron transport to the back contact, and thus considerably improves the charge-separation efficiency in a α-Fe$_2$O$_3$ NF photoanode. The α-Fe$_2$O$_3$ NF photoanode yields a photocurrent density of 3.1 mA cm$^{-2}$ at 1.5 V$_{RHE}$, and a remarkably low onset potential of 0.5-0.6 V$_{RHE}$ under AM 1.5G simulated sunlight (100 mW

cm$^{-2}$) in 1 M KOH. The lower onset value in this work is the lowest report so far combined with one of the highest photocurrent for PEC water oxidation observed for a 1D α-Fe$_2$O$_3$ photoanode without additional doping element up to now. We believe this work not only represents a significant step towards an increased photoresponse for hematite photoanodes, but also introduces a new strategy for designing 1D efficient photoanodes.

**Experimental Section**

For the preparation of α-Fe$_2$O$_3$ NFs on Fe, we used iron foils (Alfa Aesar, 99.99%) that were degreased by sonicating in acetone and ethanol for 10 minutes, followed by drying in a nitrogen stream. The samples then were thermally annealed in a furnace (Heraeus Furnace, ZEW 1451-4) in air at 400 °C. For this, the samples were placed in a ceramic boat and inserted in the furnace at room temperature. The temperature was ramped up with a heating rate of 10 °C min$^{-1}$, kept at the desired temperature for 1, 2, and 3 h, and finally the samples were removed from the furnace.

For the preparation of α-Fe$_2$O$_3$ NFs on Au/Fe, the iron foils were immersed in 5 mM HAuCl4 (Sigma-Aldrich, 99.9%) aqueous solutions for various times (10 sec, 1 min, 30 min, and 60 min) by in situ precipitation of Au nanoparticles on the top surface of Fe foils, following by drying in an air stream. Then, the Au/Fe foils were thermally annealed in the furnace in air at 400 °C, with ramping rate of 10 °C min$^{-1}$, kept at the desired temperature for various times, and finally the samples were removed from the furnace.

For the preparation of α-Fe$_2$O$_3$/FeOOH NFs on Au/Fe, the 5 mM FeCl$_3$ (Sigma-Aldrich, 99%) aqueous solutions were prepared. The samples were immersed in the FeCl$_3$ solutions for various times by a facile "top-down" method (The front sides of NFs were face to the bottom), following by rinsing with distilled water and drying in a nitrogen stream.

Characterization. X-ray diffraction (X'pert Philips MPD with a Panalytical X'celerator detector, Germany) was carried out using graphite monochromized Cu Kα radiation (Wavelength 1.54056 Å). A field-emission scanning electrode microscope (Hitachi FE-SEM S4800, Japan) was used for the morphological characterization of the electrodes. TEM analysis was performed on a FEI Tecnai TF20 microscope operated at 200 kV. Raman spectra were acquired using a Renishaw in Via Reflex Confocal Raman Microscope with an excitation laser wavelength of 532 nm. UV-vis diffuse reflectance spectra were taken on an UV-2550 (Shimadzu) spectrometer by using BaSO4 as the reference. The Fourier transform infrared spectra of the samples were recorded using a Nexus 870 spectrophotometer from 4000 to 500 cm$^{-1}$ equipped with an ATR assembly.

Synchronous illumination X-ray photoelectron spectroscopy (SIXPS). System diagram of the synchronous illumination X-ray photoelectron spectroscopy (SIXPS) was shown in Figure 3d. Different from the normal XPS instrument (PHI 5600, spectrometer, USA), a 300 W Xe arc lamp equipped with a cut-off filter (UVREF400) as illumination source is adopted. The changes of XPS spectra were recorded by controlled light on/off at given time intervals during the measurement.

Photoelectrochemical performance. The photoelectrochemical experiments were carried out under simulated AM 1.5 (100 mW cm$^{-2}$) illumination provided by a solar simulator (300 W Xe with optical filter, Solarlight; RT). The 1 M KOH aqueous solution was used as an electrolyte after saturation with N$_2$ gas for 30 min. A three-electrode configuration was used in the measurement, with the α-Fe$_2$O$_3$ electrode serving as the working electrode (photoanode), an Ag/AgCl (3 M KCl) as the reference electrode, and a platinum foil as the counter electrode. Photocurrent vs. voltage (I-V) characteristics were recorded by scanning the potential from -0.7 to 0.7 V (vs. Ag/AgCl (3 M KCl)) with a scan rate of 10 mV s$^{-1}$ using a Jaissle IMP 88 PC potentiostat. The measured potentials vs. Ag/AgCl (3 M KCl) were converted to the reversible hydrogen electrode (RHE) scale using the relationship $E_{RHE}=E_{Ag/AgCl} + 0.059 pH + E^0_{Ag/AgCl}$, where $E_{Ag/AgCl}$ is the experimentally measured potential and $E^0_{Ag/AgCl}$ = 0.209 V at 25 °C for an Ag/AgCl electrode in 3 M KCl. The stability measurement was tested at 0.5 V (vs. Ag/AgCl (3 M KCl)) in 1 M KOH under AM 1.5G illumination. Gas chromatography (Agilent; GC-7890A, MS-5A column, TCD, Ar carrier) was used to analyze the evolved gases. Photocurrent spectra were acquired at an applied potential of 0.5 V (vs. Ag/AgCl (3 M KCl)) in 1 M KOH recorded with 20 nm steps in the range of 300-700 nm using an Oriel 6365 150 W Xe-lamp equipped with an Oriel Cornerstone 7400 1/8 m monochromator. Electrochemical impedance spectroscopy (EIS) measurements were performed using a Zahner IM6 (Zahner Elektrik, Kronach, Germany). EIS measurements were performed by applying 1.5 VRHE at a frequency range of 105 Hz to 0.01 Hz with an amplitude of 10 mV in the light condition (525 nm (150 mW) laser point). Intensity modulated photocurrent spectroscopy (IMPS) measurements were carried out using Yahner IM6 (Zahner Elektrik, Kronach Germany) with a visible light (λ=429 nm) in 1 M KOH.

**Acknowledgements**

The work was supported by the National Natural Science Foundation of China (21622310, 21573264, 21633013), and the ERC, the DFG and the DFG cluster of excellence "Engineering of Advanced Materials" (EAM).

**Keywords**: single crystalline hematite photoanodes • Au layer • ultra-thin FeOOH cocatalyst • water splitting

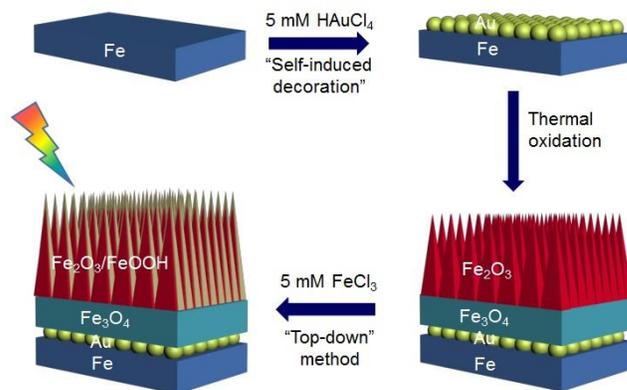

**Scheme 1**. Schematic diagram for the formation of α-Fe2O3/FeOOH NFs on Au/Fe.

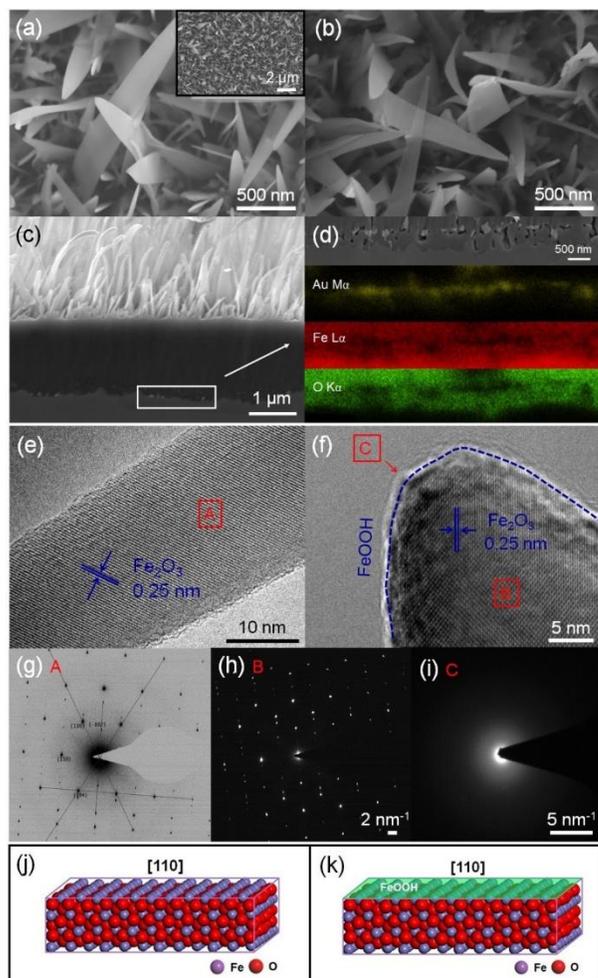

**Figure 1.** (a-c) SEM top and cross sectional images of α-Fe2O3 NFs on Au/Fe (a,c) without and (b) with FeOOH decoration; (d) representative SEM-ED mapping of α-Fe2O3 NFs on Au/Fe; (e,f) TEM images and (g-i) SAD patterns of α-Fe2O3 NFs on Au/Fe (e,g) without and (f,h,i) with FeOOH decoration; (j,k) atomic construction model of (j) α-Fe2O3 NFs on Au/Fe and (k) α-Fe2O3/FeOOH NFs on Au/Fe.

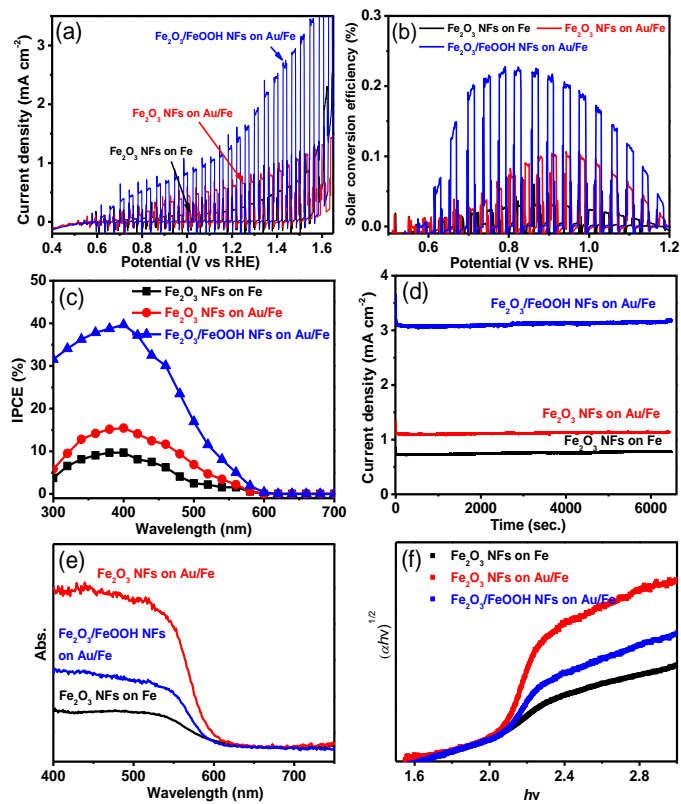

**Figure 2.** (a) PEC water splitting properties of α-Fe2O3 NFs on Fe, α-Fe2O3 NFs on Au/Fe, and α-Fe2O3/FeOOH NFs on Au/Fe. The experiments were carried out in 1 M KOH electrolyte under AM 1.5G (100 mW cm-2) illumination; (b) solar energy efficiency of α-Fe2O3 NFs on Fe, α-Fe2O3 NFs on Au/Fe, and α-Fe2O3/FeOOH NFs on Au/Fe; (c) incident photon conversion efficiencies (IPCEs), (d) stabilities, and (e,f) UV-vis diffuse reflectance spectra, and calculation of band gaps of by Tauc plots of α-Fe2O3 NFs on Fe, α-Fe2O3 NFs on Au/Fe, and α-Fe2O3/FeOOH NFs on Au/Fe. Stabilities are measured in 1 M KOH electrolyte at 1.5 $V_{RHE}$ under AM 1.5G (100 mW cm-2) illumination.

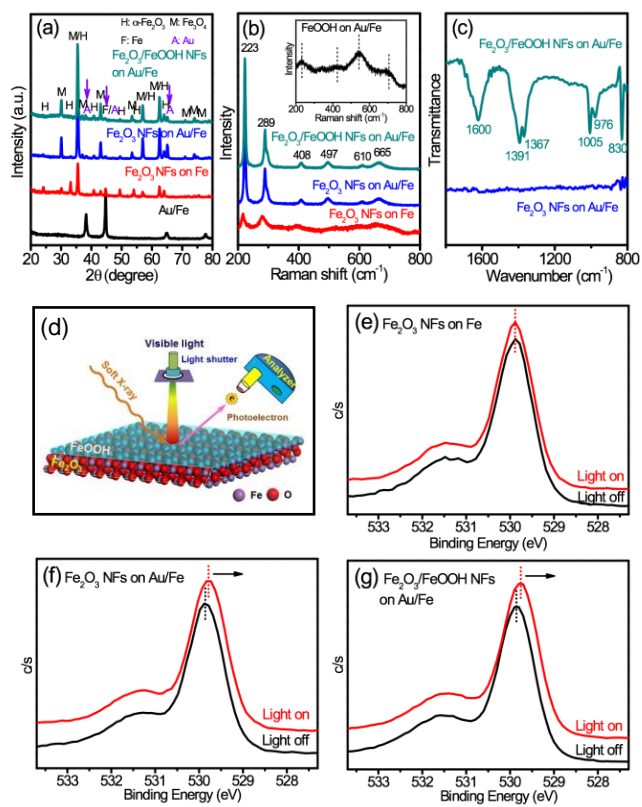

**Figure 3.** (a) XRD patterns, (b) Raman spectra, and (c) FTIR of α-Fe2O3 NFs on Fe, α-Fe2O3 NFs on Au/Fe, and α-Fe2O3/FeOOH NFs on Au/Fe. Inset of b shows Raman spectroscopy of FeOOH on Au/Fe; (d) schematic diagram of synchronous illumination XPS (SIXPS) technique; (e-g) SIXPS spectra high resolution O 1s spectra for (e) α-Fe2O3 NFs on Fe, (f) α-Fe2O3 NFs on Au/Fe, and (g) α-Fe2O3/FeOOH NFs on Au/Fe.

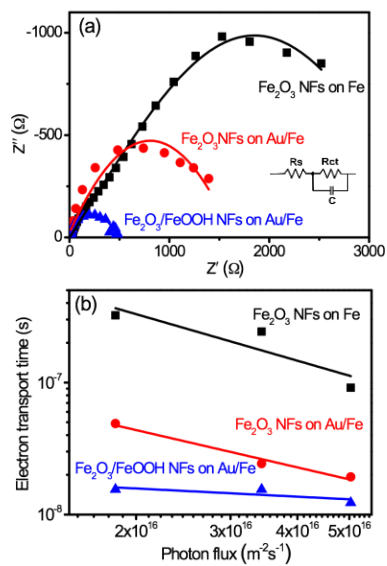

**Figure 4.** (a) Electrochemical impedance spectroscopy and (b) intensity modulated photocurrent spectra (IMPS) characterization of α-Fe2O3 NFs on Fe, α-Fe2O3 NFs on Au/Fe, and α-Fe2O3/FeOOH NFs on Au/Fe. The EIS measurement was carried out in 1 M KOH at 1.5 VRHE. Inset of (a) shows an equivalent circuit model for the photoanodes. Rs represents the solution resistance; the capacitance Cct and resistance Rct characterize the charge-transfer behavior across the electrode–solution interface; a light-emitting diode (λ=429 nm) was used as the modulation light source.